# Meeting the SDGs: Enabling the Goals by Cooperation with Crowd using a Conversational AI Platform

Jawad Haqbeen
Department of Computer Science
Nagoya Institute of Technology
Nagoya, Japan
jawad.haqbeen@itolab.nitech.ac.jp

Takayuki Ito
Department of Social Informatics
Kyoto University
Kyoto, Japan
ito@i.kyoto-u.ac.jp

Sofia Sahab
Department of Computer Science
Nagoya Institute of Technology
Nagoya, Japan
sahab.sofia@nitech.ac.jp

Rafik Hadfi
Department of Computer Science
Nagoya Institute of Technology
Nagoya, Japan
rafik.hadfi@nitech.ac.jp

Takumi Sato
Department of Computer Science
Nagoya Institute of Technology
Nagoya, Japan
sato.takumi@itolab.nitech.ac.jp

Shun Okuhara
Department of Computer Science
Nagoya Institute of Technology
Nagoya, Japan
okuhara.shun@nitech.ac.jp

*Abstract*—In this paper, we report about a large-scale online discussion with 1099 citizens on the Afghanistan Sustainable Development Goals (A-SDGs), a localization of the SDGs for Afghanistan, conducted on an AI-enabled web discussion platform utilized for gathering public opinions. We methodically used social platforms to redirect participants into the AI-enabled discussion room for insight extraction in collaboration with Ministry of Economy and Kabul Municipality, Afghanistan. The goal of experiment was to increase the inclusion of citizens' opinions in the implementation of the SDGs. Additionally, we clarified the necessity of prioritizing public intention-based SDGs goals via the driving of insights for each goal using AI and social platforms in collective intelligence at large. Furthermore, we clarified that participation was promoted by leveraging social platform functions within our system and engagement with the discussion promoted by the efficiency of AI-based discussion mediation, facilitation, summarization and visualization operations in real-time. We presented statistical information about a total of 5104 opinions provided by 1099 participants through both AI insights identification and human-led study. We found that in discussion with agent's facilitation support, the agent improved the responsiveness of citizen and increased the number of IBIS's ideas and pros compared to the issues and cons.

## I. Introduction

In order for municipalities to localize the Sustainable Development Goals (SDGs), they must employ intelligent systems to proactively promote public education and engagement. The fundamental nature of SDGs' local focal points can allow municipal governments to create means for mobilizing citizens to become active stakeholders and participants, via collective intelligence gathering, in the adaptation of the SDGs and the cities' ecosystems. The way to address the localization issues of SDGs is not monolithic. Instead, it should be centered on individual citizens' small contributions that, through collective intelligence, allow municipal government to work with society towards a sustainable city and to improve people's quality-of-life. Thus, to achieve the SDGs, we must evolve from an individual real-site discussion system to an AI-enabled cloud-based crowd discussion system, i.e., a discussion platform where everyone, everywhere and anytime can contribute by submitting their opinions via collective intelligence.

Web discussion is considered the next generation and the most important smart democratic discussion tool in order to promote equal participation and engagement. This is true both for town related meetings for the good of open society and for cooperation in achieving the SDGs. These forums, drawing on both crowd collaboration and Artificial Intelligence, can play an important role towards promoting citizen oriented urban-related development policies. The AI-enabled web forums in particular ensure the inclusion of citizens' opinions about the challenges they face by facilitating and mediating the discussion.

The Urban forum can be enhanced by employing Artificial Intelligence (AI) to facilitate the people's cooperation at large with the SDGs. In this case, conversation AI is required to be effective to perform discussion facilitation and insight extraction for large-scale discussion. Also, Conversational AI platforms can play an important role in converging diverse stakeholders' forums, such as social, executive, administrative and political forums and networks. The convergence of different level forums from low to high level, and from social to political and executive to administrative can help to collaboratively achieve the SDGs. To this end, much attention has been focused on Conversational AI systems as an important tool to promote public participation and deliberate the discussion, and the localization of the SDGs in an open and transparent manner.

Local governments are key actors for the achievement of SDGs, not only because they deliver essential services, but also for their capacity to mobilize local stakeholders and civilians. Therefore, promoting civilian engagement and enabling local stakeholders to contribute to the localization and implementation of the SDGs is key for their achievement.

Furthermore, effective facilitation should be provided to civilian and local stakeholders throughout the discussion. Insights need to be identified and mined by using smart tools such as AI technologies for a more sustainable localization of the SDGs and policy-making. In other words, the linking of public insights collected in different fields is effective and required for sustainable localization of the SDGs.

This work was supported by JST CREST fund (Grant No. JPMJCR15E1) Japan.

Therefore, in this work, we conducted a large-scale social experiment considering the Afghanistan country level SDGs as the focal point through the Ministry of Economy and in collaboration with Kabul municipality. For this, we used a Conversational AI System. Additionally, we leveraged the strength of social network platforms, such as Facebook, by redirecting the social flow into our system discussion room to converge discussion on SDGs cooperation. Our system contributed to the practical addressing of the SDGs as they pertain to Afghanistan, particularly challenges and opportunities in the localization and implementation, through promoting discussion and gathering large-scale public insight and complementing the process through the utilization of Conversational AI. This allowed for the second study: the effect of AI on promoting the submission of and collecting of people's opinions.

This project was the first time a large-scale web discussion was organized on SDGs in Afghanistan, and the first time to collect public opinion with the ability to identify ideas and issues discussed in real-time. It was, of course, therefore also the first time that social media platforms such as Twitter and Facebook were utilized to such an end.

The rest of the paper is structured as following. In Section 2, we cover literature relative to enabling SDGs by using AI and its effect to help realize the fulfilment of the SDGs. In Section 3, we present our research methodology. In Section 4, we demonstrated the system utilized. In Section 5, we employ crowd in collaboration with A-SDGs, with the support of social platforms using conversational AI to conduct a social experiment as our key-method of data collection, and present the social and experimental setting. In Section 6, we present the result of the social experiment. Finally, we present the discussion in Section 7, and summarize our finding and provide future research directions in Section 8.

## II. BACKGROUND AND RELATED WORK

Discussion platforms are considered to be the next-generation democratic platforms for citizen deliberation in collective intelligence [1]. For instance, the CoLab platform was used to harnesses the collective intelligence of thousands of people all around the world to meet the goals related to global climate change [2]. A large-scale web discussion support system, "Collagree", was used to conduct a large-scale social experiment for gathering citizen's opinions for the next-generation city planning of Nagoya city, Japan [3]. The MIT "Deliberatorium" was created and used to enable large-scale deliberation about complex systemic problems [4]. Such AI-enabled web platforms are also used to empower citizens to achieve country-specific sustainable goals [5]. For instance, An AI-assisted deliberative web discussion was employed to collect citizens' opinions for the Kabul city decade of action plan policy-making program in Afghanistan [6-7]. Another project has recently used a web platform as an AI representative discussion application to fight COVID-19 by collecting and analyzing vast amounts of social data regarding collected from COVID-19 frontline fighters, patients and private citizens using AI web discussion to increase public awareness through visualizing collective intelligence discussion trees of each group's concerns in real-time during the pandemic era [9]. AI-based facilitation is poised to play an increasing role in generating the opinions and the effects of agent-based facilitation has been verified by authors of [19-21]. The incentive mechanism in such a web discussion platform are also used to motivate citizens to participate in discussion. For instance, the authors in [22] designed an incentive mechanism for their state-of-the-art discussion platform to manage incentivizing large-scale web-based discussions. Additionally, these types of scoring mechanism help to outline the quality of submitted opinions as expressed by participants [26]. Many researchers have focused on facilitator and incentive mechanism roles that can manage inflammatory language and encourage positive discussion by utilizing such web-based discussion system. However, there is a problem that the number of posts and views decrease with a time lapse if we set discussion session for a participant to be able to comment on discussion at any hour, 7 days a week. In [23] study, the authors proposed a core-time mechanism to manage internet-based discussion. The core time mechanism provides setting of time for facilitator and participants to gather and discuss in specific time. By presenting the core time to the participants, there is expectation to encourage the participation of discussion at that time. These types of system can also help the municipal circles to host the town-related workshop and facilitate among participants to reach an agreement. For instance, a study verified the effect of a consensus building support platform "COLLAGREE" in continues municipal workshop for city development [24]. They verified that this type of system can be very useful to host pre and post-real-site municipal workshop using such web-based system to enable the city to collect divergence opinions along with its conclusion. In another study, the authors verified that using a web-based discussion enhances participation and stratification among audience in the floor-based panel discussion [25]. After all, it is very important as a human experimenter to use methods and techniques to analysis the collected discussion data to understand the insights behavior. The authors in [27], proposed a method to design an effective web-based discussion to collect participants opinions effectively, and then, analysis those collected data from internet-based discussion. Their aim was to design a discussion to collect discussion data effectively as a training data of deep learning for the development of automated facilitation system.

In practice, these AI-assisted discussion platforms employ algorithmic methods and machine learning techniques to harness the intelligence of the crowd. In this work, we particularly focus on the use of conversational agent to adopt human conversational behavior and provide support by facilitated messages to lead the discussion on the SDGs. In this case a defined computer program as a representing AI facilitator interacted with users using natural language processing technologies to mimic discussant-submitted items throughout the conversation on a large-scale. However, using an AI-enabler in discussion platforms raises ethical problem and has ethical side effects on the discussion. Specifically, this is because the agents are expected to exercise judgment when interacting with discussants, and has shown to have great capability for changing the minds of discussants through argumentative messages. For instance, several studies have investigated the effect of artificial agents on social media platforms. In [10], the work focus on the polarization effects of an agent's activities on a political social network by studying the retweet network of 3.7 million users during the Stoneman Douglas high school shooting event. The research found that agent accounts heavily contributed to online polarization. In our work, we conducted a non-polarized discussion where the conversational agent simply provided participants with support throughout discussions via mediation, as opposed to raising counter-ideas in an attacking manner. We set a 3:1 (participant post: AI post) ratio for agent interaction. This means the agent provided a facilitating message once after every three participant posts and did not post to all participants. In the future, we will study how those who received posts from the agent and those who did not may have been differently affected. Also, we will consider the side effects of the agent by studying that how such agents can help to increase the percentage of polarization on items discussed and in the discussions in general.

## III. METHODOLOGY

The general methodology of our research centers on conducting a discussion between human participants with a conversational agent as a facilitator, wherein the agent is expected to encourage participants to provide the ideas on issues related to localizing the A-SDGs. We distinguish the effects of AI by not employing the agent in the first month of discussion. We also tried to use soft system methodology in collaboration with crowd (A-SDGs) using conversational AI to identify the effects of the AI web platform raising elements of discussion. Our methodology benefited from computational tools, such as agent-based simulation, machine leering, mining technology, deep learning. It further benefited from crowd collaboration such as in the case of the joint project with Kabul Municipality, Afghanistan.

### A. Discusssion with Crowd

This study used the purposed AI-enabled discussion system (D-Agree) as an AI representative deliberative discussion application to identify and summarize the public's SDG- related responses. The system and social platform were methodically used to collect social network opinions; we redirected the social flow from Facebook by boosting Facebook Ad. service into our system. Via Facebook Ad., we reached out to our target the audience in all of Afghanistan's provinces.

As such, people all over the nation and around the world were able to share their thoughts through a large-scale discussion support system on how to achieve the SDGs in Afghanistan. The call for participation was posted on Kabul Municipality's and Afghanistan's SDGs country focal point's respective official Facebook and Twitter accounts and homepages. The local government also asked residents to register for the social experiment. We received 1099 registrants, of which 82 were female. The experiment was conducted from January 20th to May 25th in two phases: a one-month discussion without an agent and a 3-month discussion with an agent. We evaluated the effects of the agent by comparing two equal time periods, namely the first month of agent-free discussion and the subsequent debut month of the agent. The participants were able to comment on discussion at any hour, 7 days a week. Discussion revolved around the five following themes:

1. How A-SDGs should be adopted effectively
2. How to promote civic engagement in A-SDGs
3. The key elements that may need to be taken into account in formulating roadmap for the localization of and realization of the SDGs
4. The ranking, in terms of prioritization, of SDGs in regards to the application of the A-SDGs
5. What a successful forum to engage citizens on the SDGs discussion would look like

There were two objectives behind the crowd discussion. First, the Kabul municipality and the Ministry of Economy, the country's focal point for the SDGs in Afghanistan (A-SDGs), wanted to promote public engagement by collecting insights from citizens for country-related SDGs. Second, we wanted to verify the effect of an agent in non-polarized discussions in a developing country by conducting large-scale social experiment with and without said agent.

### B. Non-Polarized Discussion without Coversational Agent

The objective behind this discussion was to conduct a social discussion without introducing an agent to check how the responsiveness of participants leading the discussions and how the discussion tree would be structured without a facilitator in the discourse. We chose not to activate the agent in the first month of discussion and to open discussions with humans alone to allow the agent to learn from the discourse mined data and then to compare the characteristics of participants. Characteristics included such issues as discussion responsiveness and those related to Issue Based Information System (IBIS) elements. This allowed for the study of the effects of increasing each element within both discussions. The experiment was conducted from January 20 to February 16, 2020. We selected the first two themes from a theme list for discussion without an agent.

### C. Non-Polarized Discussion with Coversational Agent

The objective behind this discussion was to introduce an agent to check the responsiveness of participants leading the discussion and how the discussion tree will be structured with a facilitator with a discourse facilitated by an agent. Also, we wanted to test the effects of AI facilitation in leading the discussions and raising the discussion elements. The experiment was conducted from January 17 to May 28, 2020. We looked at the second month of discourse for comparison with discussion without the agent. We selected all five discussion themes but considered first to compare with discussion without the agent.

## IV. OUR SYSTEM DEMO

### A. Web-based AI Discussion System (D-Agree)

Our team created and used D-Agree, an online AI-enabled discussion system for hosting large-scale discussions and real social experiments [3, 6-9]. Through this system, participants can submit their opinions as text. D-Agree is the upgraded version of CollAgree, also developed by our team. The main difference between CollAgree and D-Agree is the server management side; whereas in CollAgree is based on a local computer, D-Agree is on the cloud. We created and used D-Agree [1-5] to promote engagement and facilitate the discussion via collective intelligence. The system can be used by anyone without installing the app. It is supported on both mobile and desktop. Also, the app is available on the Google Play store for android users. The current version is available in Japanese and English. The two main agent technologies used while designing the mechanism of our software agent are the real-time text extraction of discussion, visualization of insights, and autonomous posting of facilitation messages by the AI agent to lead discussion towards general consensus-building. In other words, our agent observes the posted texts, extracts their semantic discussion structure, generates facilitation to lead discussions and, by underling the bold insights, helps crowds to reach an agreement. This is all active 24 hours a day, 7 days a week.

The system adopted the IBIS model as a discussion framework [11]. It employs and extracts users' posted items and for each user-posted data, a set of features is learned automatically via a discussion structure module. After that, classification will be applied based on the IBIS structure nature. Then, output will be visualized to predicate posted item traits and behaviors as issues, ideas, pros and cons. The main reason to adopt IBIS was to lead the extension of argumentative the discussion structure through which people clarify issues, ideas, merits and demerits as pertains to the SDGs promotion [6]. Figure 1 presents the discussion structure of our system.

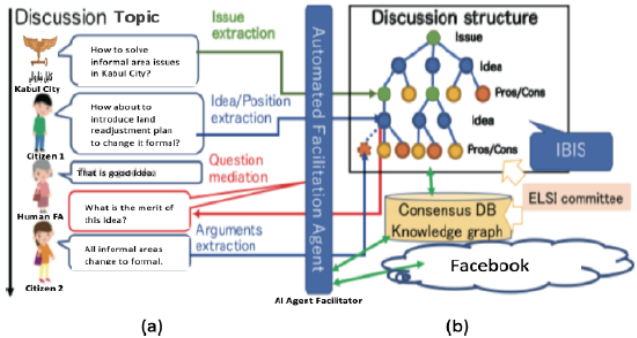

Fig.1: (a) discussion structure extractions, (b) discussion structure, adopted from IBIS.

### B. System Architecute and User Interface

We used Amazon Web Services to maintain the scalability of large-scale discussion operation. Fig. 3 presents the steps of social login to the system and Fig. 3(f) present user interface of our system in which we have posting, thread, theme, and point areas. The post form is used to input an opinion and for collecting people's opinions. Once the discussion topic has been posted, a user can reply to the topic's post form or submit opinions under the topic. In this way, the discussion leads to structure and creates a thread area. Also, a user can select the satisfaction level from 1-10 and input his/her opinion in a topic's post form or submitted opinions. The range from 1-5 represent opposing views and 6-10 shows levels of agreement. Theme is the top-level issue and is usually defined by an administrator while creating the discussion. Fig.2 present the system general architecture in which we developed a general sense automated facilitation agent that observes user-generated data, extracts their semantic discussion structures, generates facilitation messages, preserves and maintains the health of discussion by filtering out inappropriate data, classifies the posted items and visualizes the categorization of debaters' posted items.

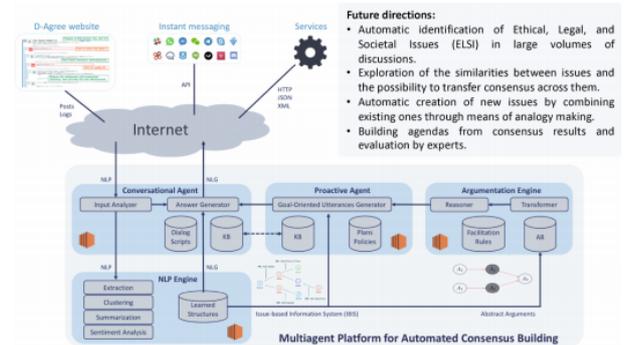

Fig. 2. System general architecture

Two main mechanisms have been considered while designing the agent: a discussion extraction/visualization mechanism and an observing and posting mechanism. The social login system is shown in figure 3.

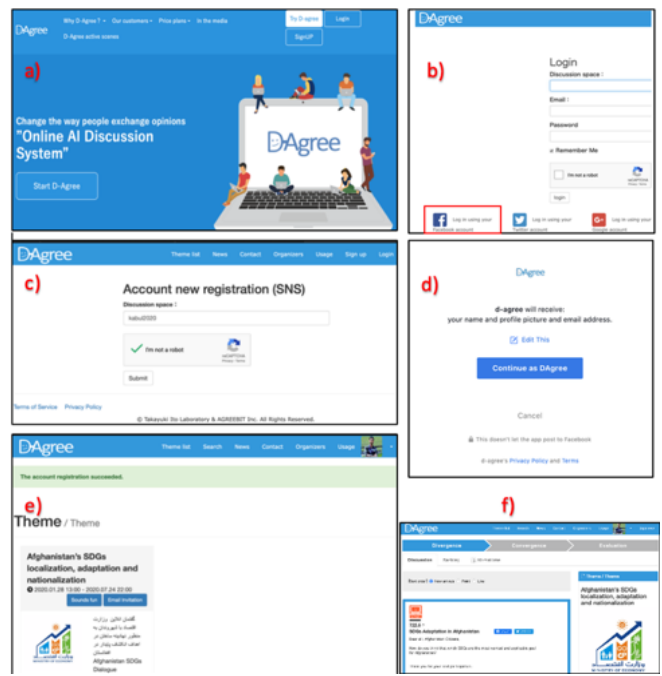

Fig. 3. Social login demo of our system

### C. Link and Node Extractor

Real-time analysis of the discussion is performed in D-Agree. This technology can be said to be the most significant feature of D-Agree and it allows flexible automatic facilitation, unlike chatbot, and discussion summarization system implementation. More specifically, we are extracting the structure of the discussion in real-time. These extractions include node extraction and link extraction [12]. The node extraction automatically classifies the sentences in the discussion into four classes: "issue", "idea", "pros", "cons". The link extraction predicts a relationship between sentences. These four classes are based on the Issue-Based Information Systems (IBIS) structure proposed by Kunz et al [11]. The IBIS structure is shown in Fig. 4. "Issues" is for problems to solve, "Ideas" is for proposed solutions to a problem, "Pros" is for the advantages of the idea, and "Cons" is for the shortcomings of the idea. Each of these elements is called a node, and a relationship between the elements is called a link. The automated facilitation agent uses the IBIS structure to manage the discussion. For example, the agent facilitates suitable content for the extracted nodes. Also, the D-Agree discussion summarization system uses the extracted node and link information. This system reconstructs the discussion according to the IBIS structure and displays for users.

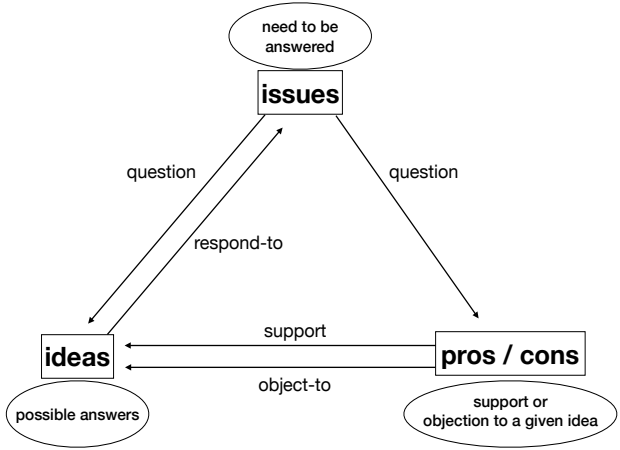

Fig. 4. Outline of IBIS structure and Extraction flow

Table 1. (a) Results of Node Extraction and (b) Link Extraction

(a)

| IBIS | Precision | Recall | F-measure |
|---|---|---|---|
| Issue | 0.89 | 0.89 | 0.89 |
| Idea | 0.84 | 0.89 | 0.87 |
| Pros | 0.86 | 0.79 | 0.82 |
| Cons | 0.82 | 0.79 | 0.80 |

(b)

| | Idea --> Issue | Pros --> Idea | Cons --> Idea |
|---|---|---|---|
| Precision | 0.488 | 0.235 | 0.209 |

The node extraction and the link extraction are based on a deep learning method. As training data, we use our original dataset of discussions conducted using D-Agree in English. These discussions were on 27 different topics, such as social issues and concerns of foreigners living in Japan (translators, English teachers, and students in Nagoya Institute of Technology). Each topic was discussed by four people. 2050 Data sets were used for the training of the node extraction and 955 Data sets were used for the training of the link extraction. Annotation was done manually. We used BERT [13] and Dense as a model of node extraction and used fastText [14] and Bi-LSTM [15] as a model of link extraction. The evaluation results of node extraction by using 3-fold cross-validation are shown in Table 1, and the evaluation results of link extraction by using leave-one-out cross-validation [16] are shown in Table 2. In the result of the link extraction, we did not evaluate "issue --> idea" and "issue --> pros/cons" because there was not much annotated data. Currently, to improve accuracy, we have proposed a new method for node extraction [17]. Using the Graph Attention Network (GAT) [18], it can train not only sentences but also graph structures. The method using GAT has shown good results, and we are working on its implementation in the D-Agree.

V. SOCIAL EXPERIMENTAL SETTING

A. Collaboration Setting

In September 2019, we had an opportunity to establish collaboration on a series of large-scale social experiments with Kabul Municipality. The main purpose was to promote urban related dialogue by using a large-scale discussion support system. We proceeded to expand the collaboration with the Ministry of Economy, which is the focal point of SDGs in Afghanistan, to check how crowd's collaboration can be harnessed using an AI-enabled web discussion platform.

B. Experiment Setting of discussion with agent

We introduced the agent to play the role of a moderator, through the posting of supportive facilitated messages, and a supportive role to raise ideas for the given issue. For example, a supporting facilitated message on an issue might be written as "Please feel free to provide anything that comes to your mind about {name}'s {issue}." Here the variable, {name} is the name of the author of the post, and {issue} is the issue extracted by agent from the posted message. Other IBIS elements such as pros or cons also mined from the discussants' messages and were exploited in the same way. The interaction between participants and the agent was controlled with two parameters: a period of 1-minute specific to Amazon CloudWatch, and a threshold of 2 people. This threshold set the number of messages that the agent should count before taking part in the discussion. For instance, in figure 2, the agent (AI Facilitator) waited for 3 messages from participants before posting her message. The agent's identity was disclosed to the participants up until the end of the experiment.

C. Collection of Attribute

The registration social form required users to sign up for system. By default, our system can collect four types of participant's attribute, namely 1) name [login user]; 2) gender [f/m/other]; 3) ID [email address]; 4) profile photo [uploaded by user/social media profile. Additionally, we can estimate 3-5 more attributes from the social login feature and data they generated on social media. It is worth mentioning that we get the participants' consent at sign up to use their encrypted data and information for research purposes. Our system provides advanced security for important discussion data and information of participants. Not only is communication encrypted, but data is also always encrypted and saved. These social attributes were very important to analyze social behavior with posted items of participants and deep linguistic analysis on conversational data was also necessary.

It is worth mentioning that over 90% of social data was collected by our system using social platforms such as Facebook, and about 10% of participants came from other networks using their email accounts, such as google mail, for registration.

## VI. EXPERIMENTAL RESULTS

The results of the whole discussion experiment are shown in figure 5. In the whole of discussions, the number of ideas and pros are higher than the number of issues of cons. In the experiment, the number of identified opinions classified as *issues* were 1833 items, *ideas* 1862 items, *pros* 756 items and *cons* 653 items. There was only one agent as a facilitator. In discussion experiments without an agent, the number of issues and cons were higher than in discussions with an agent. Moreover, in discussions with an agent, the ideas and pros were higher than in discussions without an agent. In figure 6, the blue color represents the outline of submitted opinions in discussions without an agent and orange outlines the submitted opinions in discussion with an agent; in total this represents the first two months of discourse.

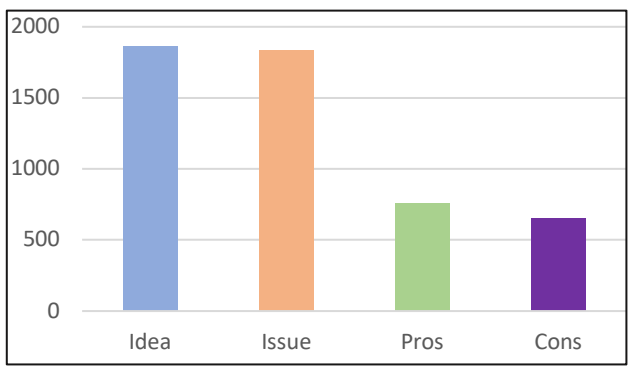

Fig. 5. Extracted IBIS from all submitted opinions

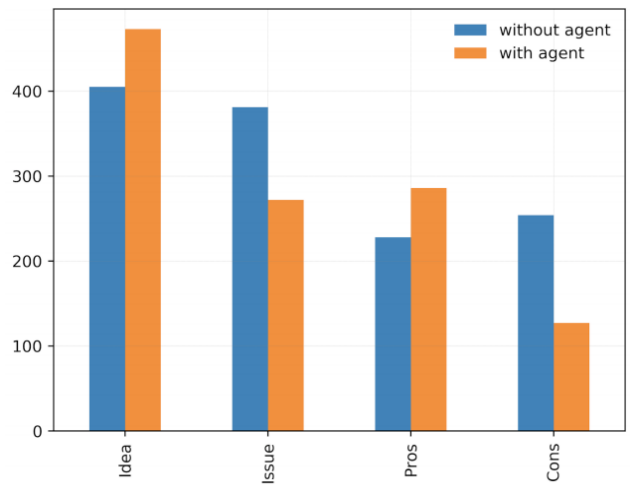

Fig.6. Extracted IBIS in discussion without and with agent

## VII. DISCUSSION

Our goal was to take the opportunity of crowd collaboration, and probe into it using both social platforms and an AI-based discussion system for a collective intelligence approach to promote awareness and collect insights related to SDGs in collaboration with governmental organizations in Afghanistan. First, the IBIS-based discussion trees have been structured from participant-submitted opinions to check how the SDGs are being met and compare crowd discussion behavior. The evolution of these discussion tree elements suggests that AI-assisted web discussion can promote the discussion towards reaching general consensus by checking the conversational agent effect on evolving discussion to predominate discussion elements, which can lead discussion towards a positive conclusion. We need further study to verify the accuracy of the agent towards both positive and negative side effects throughout this discourse. Also, we have to reconfirm the accuracy of predomination of IBIS elements by the agent. To do this, we need to verify the effect of IBIS predomination on a large-scale group of people with different cultures and education backgrounds. We have noticed some constraints during our social experiment. Also, as most studies based on social platform are biased, we also confirmed the presence of bias in data collection. Having said that, the goal of this large-scale experiment on SDGs was to provide an insightful view at large through aggregating public opinions by using AI. Due to nature of this study, and considering these facts, this is the first time we using conversational AI to conduct large-scale discussion and aggregating public opinion to present deep insights for localization of SDGs in Afghanistan. Also, we are confident that the results presented here and large-scale social insights which were presented to Ministry of Economy and Kabul Municipality will help in appreciating the sequence of events that transpired and better prepare sustainable goals.

### A. Gaps in Opinion Collection

Due to lack of popularity in Afghanistan social circles of the system used, and the fact that the new system is not a primary social discussion tool for people in Afghanistan, such as Facebook and Twitter, we were not able to reach 10K-scale. We initially did take this possibility into account by calling for participation using different real and cyber resources as quickly as we conceived of this project, but this did not overcome the gap. Additionally, due to high prices for internet access, people usually buy specific internet packages, such as a Facebook package, so that their installed internet service coverage can be limited only to social media platforms and they cannot login beyond them to other services/sites. It is likely that we have inadvertently missed out on reaching many participants due to this. We reached to 141,312 people in total by boosting social ads using Facebook but we collected only 1099 participants for our experiment. Furthermore, the challenge we faced was the technology literacy and the cultural fact that people in Afghanistan, even literate ones, are not familiar with such a discussion system.

VIII. CONCLUSION

In this paper, we have reported the case study of large-scale social experimentation on the SDGs in Afghanistan using an online AI system and social platform to gather public insights. Our system identified the IBIS elements for the localization of SDGs to strengthen the plans by bringing together people to discuss the issue online.

To conclude, the AI-enabled web support discussion had strong effects on leading the discussion to generalize consensus building. In particular, we found that a conversational agent has strong side effects on predominating those elements of discussion that help reach overall agreement for the conclusion of a discourse. We observed this effect by comparing two discourses hosted by our system and found that the impact of using an agent as a facilitator in public deliberation outperformed not using agent. Also, the using of an agent in discourse was found to help increase the idea and pros element compared to issues and cons. Our future work is to extend the discussion into groups with different cultures and educational backgrounds to check the domination of IBIS elements and the role of agents within each group. Also, we will extend the linguistic analysis on participants discussion and conversational behavior to understand the real discussion tree and its element by comparing the frequencies of words used within each element.

ACKNOWLEDGMENT

We would like to thank and express our deepest appreciation to all people who participated in discussions and the social experiment. Also, we truly appreciate the mayor of Kabul, Mr. Daoud Sultazoy, ex-mayor Mr. Zaki Sarfarz and the A-SDGs country focal point leader, Dr. Mirwais Baheej, for their support and constructive comments, as well as we would like to thank to Jason Pratt, an associate professor at Yamanashi Prefectural University, Japan for his collaboration in manuscript proof reading.

REFERENCES

[1] T.W. Malone, "Superminds: The superising power of people and computers thinking togather". Little, Brow Spark, 2018.
[2] T.W. Malone and M. Klein, "Harnessing collective intelligence to address global climate change". Innovations: Technology, governance, globalization 2(3), pp. 15–26, 2007.
[3] T. Ito, Y.a Imi, T. Ito and E. Hideshima, "COLLAGREE: A faciliator-mediated largesclae consensus support system," In proceeding of the 2nd ACM Collective Intelligence Conference pages 10-12, 2014.
[4] M. Klein, "Enabling large-sale deliberation using attention-mediation metrics," Computer Supoprted cooperative work 21(4-5), pp.449-473
[5] P. Savaget, T. Chiarini and S.Evans, "Empwering political particiaption through artificail intelligence. Science and public policy 43(3), pp.369-380, 2019.
[6] J. Haqbeen, T. Ito, R. Hadfi, T. Nishida, Z. Sahab, S.Sahab, S.Roghmal and R.n Amiryar, "Promoting discusison with AI-based Facilitation: Urban Dialogue with Kabul City," In proceeding of the 8th ACM Collective Intelligence Conference, 2020.
[7] J. Haqbeen, T.Ito, R. Hadfi, S.a Sahab, T. Nishida and R. Amiryar, "Usage & application of AI-based discussion facilitation system for urban renewal in selected districts of Kabul city: Afghanistan Experimental View," In In proceeding of the 34th Japanese Society of Artificial Intelligence, 1C3OS6a01-1C3OS6a01, 2020.
[8] A. Sengoku, T. Ito, K. Takahashi, S. Shiramatsu, T.i Ito, E. Hideshima and K. Fujita, "Discussion tree for managing large-sclae internet-based discussions," In proceeding of the 5th ACM Collective Intelligence Conference., 2016.
[9] J. Haqbeen, T. Ito, S.Sahab, R.Hadfi, S.Okuhara, N.Saba, M.Hofiani and U.Baregzai, "A contribution to covid-19 prevention through crowd collaboration using conversational AI and social platforms". In proceeding of AI for social good workshop, 2020.
[10] M. Ozer, M.Y.Yildirim and H.Davulcu, " Measuring the polirization effects of bots acocunt the us gun control debate on social media". Conference'17, Washington DC, USA, 2019.
[11] W. Kunz, H.W. Rittel, "Issues as elements of information systems," Technical report, California Institute of Urban and Regional Development, University of California, (1970) CiteSeerX 10.1.1.134.1741
[12] S. Suzuki, N.Yamaguchi, T. Nishida, A.Moustafa, D.Shibara, K.Yoshino, K.Hiraishi and T.Ito, "Extraction of online discussion structures for automated facilitation agent," In In proceeding of the 33rd Japanese Society of Artificial Intelligence, pp.150-161, 2019.
[13] Y.Lui, "Fine-tune BERT for extractive summarization," arXIv preprint arXiv: 1903.10318, 2019.
[14] P.Bojanowski, "Enriching word vectors with subword information," Transectiono of the Association for computational lingusitics vol.5, pp.135-146, 2017
[15] M.Schuster and K.P. Kuldip, "Bidirectional recurrent neural networks," IEE Transections on Signal processing, 45(11), pp.2673-2681, 1997.
[16] M.Stone, "Cross validator choice and assessment of statisticla predictions," Journal of the royal statistical society, series B (Methodological) 36(2), pp.111-133, 1974.
[17] S.Suzuki, T.Ito, A.Moustafa and R.Hadfi, "A node classification approach for dynamically extracting the structures of online discusisons," In In proceeding of the 34th Japanese Society of Artificial Intelligence, 2020.
[18] P.Velickovic, G.Cucurul, A. Casanova, A.Romero, p.Lio and Y.Bengio, "Graph attention networks," arXiv preprint arXiv:1710.10903, 2017.
[19] T. Ito, S.Suzuki, N.Yamaguchi, T.Nishida, K.Hiraishi and K.Yoshino, "D-Agree:Crowd discussion support system based on automated facilitation agent," In proceeding of the 35th AAAI conference, 2020.
[20] T. Ito, D. Shibata, S. Suzuki, N. Yamaguchi, T. Nishida, K. Hiraishi and K. Yoshino, "Agent that facilitates crowd discussions," In proceeding of the 7th ACM Collective Intelligence Conference, 2019.
[21] T. Ito, "Towards agent-based large-scale discussion support system: The effect of facilitator," In proceeding of the 51st Hawaii International Conferenbce on system sciences, 2018.
[22] T. Ito, Y. Imi, M. Sato, T. Ito and E. Hideshima "Incentive mechanism for managing large-scale internet based discussions on collagree," In proceeding of the 4th ACM Collective Intelligence Conference., 2015.
[23] T. Nishida, T. Ito, T. Ito, E. Hideshima, S. Fukamachi, A. Sengoku and Y. Sugiyama, "Core time mechanism for managing large-scale internet-based discussion on COLLAGREE," In proceeding of the 2nd IEEE international conference on agents, 2017.
[24] T. Nishida, T. Ito and T. Ito, "Varification of effects using consensus building support system in continuous workshops for city development", Journal of Science and design 3(1), pp.161-168, 2018.
[25] S. Kawase, T. Ito, T. Otsuka, A. Sengoku, S. Shiramatus, T. Matsuo, T. Oishi, R. Fujita, N. Fukuta and K. Fujita, "Cyber-physical hybrid environment using a large-scale discuison system enhances audiences'

participation and statisfication in the panel discussion", The IEICE transaction on information and systems E101.D(4), pp.847-855, 2018.

[26] K. Takahashi, T.Ito, T. Itom E. Hideshima, S. Shiramatus, A. Sengoku and K. Fujita, "Incentive mechanism based on quality of opinioin for large-scale discusison support systeem," In proceeding of the 5th ACM Collective Intelligence Conference., 2016.

[27] N. Yamaguchi,T. Ito and T. Nishida "A method for online discusison design and discussion data analysis," In proceeding of the 13th international conference on knowledge, information and creativity support systems, 2018.